\documentclass[10pt,conference]{IEEEtran}
\IEEEoverridecommandlockouts

\usepackage[T1]{fontenc}
\usepackage{graphicx}
\usepackage{titlesec}
\usepackage{setspace}

\titlespacing\section{0pt}{6pt plus 1pt minus 1pt}{3pt plus 2pt minus 1pt}
\titlespacing\subsection{0pt}{5pt plus 1pt minus 1pt}{2pt plus 2pt minus 1pt}
\usepackage{amsthm,amsmath,amssymb,bm}
\usepackage{esint}
\usepackage{mathrsfs}
\usepackage{amsfonts}

\newtheorem{remark}{Remark}

\usepackage[caption=false]{subfig}
\usepackage{algorithm}
\usepackage[noend]{algpseudocode}
\usepackage{graphics}
\usepackage{epsfig}
\usepackage{cite}
\usepackage{color}
\usepackage{booktabs}
\usepackage{verbatim}
\usepackage{stfloats}
\usepackage{cases}
\usepackage{multirow}
\usepackage[inline]{enumitem}
\usepackage{ragged2e}
\usepackage{epstopdf}

\newcommand{\transpose}{\mathsf{T}}
\newcommand{\hermconj}{\mathsf{H}}
\newcommand{\E}{\mathsf{E}}
\newcommand{\trace}{\mathtt{tr}}

\begin{document}
	
\title{Rethinking Mutual Coupling \\ in Movable Antenna MIMO Systems}
\author{
    \IEEEauthorblockN{Tianyi~Liao, Wei~Guo, Jun~Qian, Shenghui~Song, Jun~Zhang,~\textit{Fellow, IEEE}, and Khaled~B.~Letaief,~\textit{Fellow, IEEE}}
    \IEEEauthorblockA{Dept.\ of ECE, The Hong Kong University of Science and Technology, Kowloon, Hong Kong}
    \IEEEauthorblockA{Emails: ty.liao@connect.ust.hk, \{eeweiguo, eejunqian, eeshsong, eejzhang, eekhaled\}@ust.hk}\vspace{-1cm}
    \thanks{This work is supported by the Hong Kong Research Grants Council (RGC) under the Area of Excellence (AoE) Scheme Grant No. AoE/E-601/22-R, and in part by the General Research Fund (Project No. 16209524) from the Hong Kong RGC.}}
\maketitle
\IEEEaftertitletext{\vspace{-1\baselineskip}}

\begin{abstract}
    Movable antenna (MA) systems have emerged as a promising technology for future wireless communication systems. The movement of antennas gives rise to mutual coupling (MC) effects, which have been previously ignored and can be exploited to enhance the capacity of multiple-input multiple-output (MIMO) systems. To this end, we first model an MA-enabled point-to-point MIMO communication system with MC effects using a circuit-theoretic framework. The capacity maximization problem is then formulated as a non-concave optimization problem and solved via a block coordinate ascent (BCA)-based algorithm. The subproblem of optimizing MA positions is challenging due to the presence of the analytically intractable MC matrices. To overcome this difficulty, we develop a trust region method (TRM)-based algorithm to optimize MA positions, wherein Sylvester equations are employed to compute the derivatives of the inverse square roots of the MC matrices. Simulation results show significant capacity gains from leveraging MC effects, primarily due to customizable MC matrices and superdirectivity.
\end{abstract} 

\begin{IEEEkeywords}
Movable antenna (MA), mutual coupling (MC), capacity, superdirectivity, MIMO, optimization.
\end{IEEEkeywords}

\IEEEpeerreviewmaketitle
\setstretch{0.94}
\section{Introduction}\label{sec:intro}
The sixth-generation (6G) wireless communication systems are envisioned to achieve high spectral efficiency, enhanced energy efficiency, and ultra-low latency~\cite{letaiefRoadmap6GAI2019,lin2024fundamentals}. Multiple-input multiple-output (MIMO) systems play significant roles in fulfilling these performance requirements by enabling spatial multiplexing and diversity. Recently, the trend of deploying hundreds or even thousands of antennas at the base station has emerged to meet the demands of 6G~\cite{luTutorialNearFieldXLMIMO2024}. However, the increasing number of antennas incurs substantial hardware complexity, power consumption, and signal processing costs.

Reconfigurable antenna systems, such as movable antennas (MAs)~\cite{maMIMOCapacityCharacterization2024} and fluid antenna systems (FASs)~\cite{wongFluidAntennaSystems2021}, have been proposed recently to address these challenges. By enabling antenna movement, MA systems can mitigate destructive multipath combinations, thereby improving system capacity~\cite{zhuModelingPerformanceAnalysis2024,liao2025joint}. Moreover, MA systems are effective in enhancing physical layer security~\cite{tangSecureMIMOCommunication2025}, sensing performance~\cite{li2025movable}, and energy efficiency~\cite{dingEnergyEfficiencyMaximization2025}. However, existing studies on MA systems typically impose minimum antenna spacing constraints of half a wavelength, under which the mutual coupling (MC) effects are neglected. This setting presents several issues:
\begin{enumerate*}
    \item MC effects still exist even when the half-wavelength spacing constraints are imposed,
    \item the degrees of freedom in antenna movement are sacrificed, and
    \item the potential benefits of exploiting MC are forfeited.
\end{enumerate*}

The MC effect is not new in communications and was investigated in the early stages of MIMO research. In~\cite{wallaceMutualCouplingMIMO2004}, the authors modeled MIMO communication systems with MC using a circuit-theoretic framework. The impact of MC was further examined in~\cite{morrisSuperdirectivityMIMOSystems2005}, where \emph{superdirectivity} was shown to achieve beam gains exceeding the number of antennas, thereby enhancing capacity. A more rigorous and comprehensive analytical framework was later developed in~\cite{ivrlacCircuitTheoryCommunication2010}, which explicitly accounted for both transmit and receive MC effects. As MIMO technology evolved, researchers gradually adopted half-wavelength antenna spacing as the standard configuration, as it reduces channel correlation and satisfies the Nyquist sampling theorem~\cite{tse2005fundamentals}. For isotropic antennas, such spacing effectively eliminates MC effects, whereas for non-isotropic antennas, MC effects become negligible. Consequently, MC is typically ignored in most existing MIMO studies.

Nevertheless, there has been a recent trend towards reincorporating MC effects into MIMO systems~\cite{mezghaniReincorporatingCircuitTheory2024}. This resurgence is primarily driven by the advent of holographic MIMO systems~\cite{pizzo2025mutual}, where antenna elements are densely packed and MC effects become pronounced. In MA systems, antenna elements can move off the half-wavelength grid, giving rise to MC effects. The potential benefits of exploiting MC in MA systems are twofold. On one hand, MC enables superdirectivity, enhancing the beamforming gain. On the other hand, antenna movement allows designable MC matrices to better match the wireless channel, thereby achieving higher capacity.

This paper focuses on exploiting MC effects in MA-enabled MIMO systems. We consider a point-to-point MIMO system and employ the circuit-based communication model~\cite{ivrlacCircuitTheoryCommunication2010} to characterize the system with MC effects. When MC is taken into account, MAs are allowed to move closer than half a wavelength. To explore the potential capacity gain brought by MC, we formulate the capacity maximization problem as a non-concave optimization problem. Then, the problem is solved via a block coordinate ascent (BCA)-based algorithm. The subproblem of optimizing MA positions is challenging due to its non-concavity and indefinite curvature of the objective function. To overcome these challenges, we develop a trust region method (TRM)-based algorithm~\cite{yuan2015recent} for MA position optimization. Moreover, the first- and second-order derivatives of the inverse square roots of the MC matrices are derived in quasi-closed form by solving Sylvester equations~\cite{higham2008functions}. Simulation results demonstrate significant capacity gains achieved by exploiting the MC effects.

\textit{Notation:} $[\mathbf{A}]_{m,n}$, $\mathbf{A}^\transpose$, $\mathbf{A}^\hermconj$, $\det(\mathbf{A})$, and $\trace(\mathbf{A})$ denote the $(m,n)$-th element, transpose, conjugate transpose, determinant, and trace of matrix $\mathbf{A}$, respectively. 
Symbols $\partial(\cdot)$, $\Re\{\cdot\}$, and $\mathsf{E}[\cdot]$ represent the partial derivative, real part, and expectation operator, respectively. $\mathcal{CN}(\mathbf{0},\sigma^2\mathbf{I})$ denotes the circularly symmetric complex Gaussian (CSCG) distribution with zero mean and covariance matrix $\sigma^2\mathbf{I}$.

\section{System Model and Problem Formulation}\label{sec:system_model}
\subsection{MA-Enabled MIMO System}
\begin{figure}[!t]
    \centering
    \includegraphics[width=0.48\textwidth]{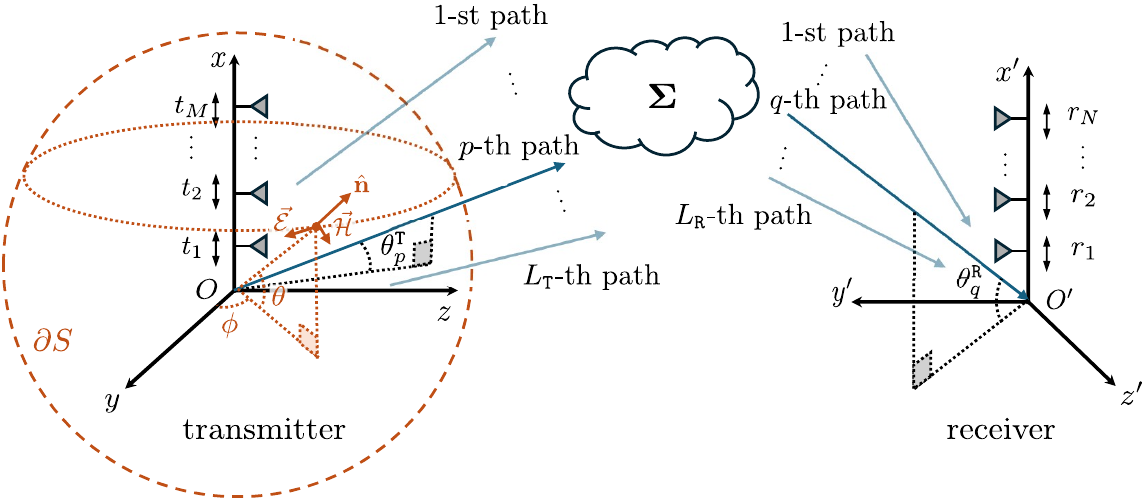}
    \caption{System model of a point-to-point MIMO communication system equipped with $M$ transmit MAs and $N$ receive MAs.}\label{fig:system_model}
    \vspace{0.1cm}
\end{figure}
As illustrated in Fig.~\ref{fig:system_model}, we consider a point-to-point MIMO communication system equipped with $M$ transmit MAs and $N$ receive MAs. Local Cartesian coordinate systems, $(x,y,z)$ and $(x^\prime,y^\prime,z^\prime)$, are established at the transmitter and receiver, respectively. The transmit and receive MAs are constrained to move along the $x$-axis and $x^\prime$-axis of their corresponding coordinate systems. Therefore, the positions of the $m$-th ($m=1,2,\dots,M$) transmit MA and the $n$-th ($n=1,2,\dots,N$) receive MA are represented by the scalars $t_m$ and $r_n$, respectively. The position vectors of all transmit and receive MAs are denoted by $\mathbf{t} = [t_1, t_2, \dots, t_M]^\transpose \in \mathbb{R}^M$ and $\mathbf{r} = [r_1, r_2, \dots, r_N]^\transpose \in \mathbb{R}^N$, respectively.

We adopt the far-field field response model~\cite{maMIMOCapacityCharacterization2024} to characterize the wireless channel between the transmitter and receiver. The transmit and receive steering vectors are defined as

\begin{small}
\vspace{-0.3cm}
\begin{subequations}\label{eq:steering_vector_def}
    \begin{align}
        \mathbf{a}_{\mathtt{T}}(\mathbf{t},\theta) = [\mathrm{e}^{j \kappa t_1 \sin\theta}, \mathrm{e}^{j \kappa t_2 \sin\theta}, \dots, \mathrm{e}^{j \kappa t_M \sin\theta}]^\transpose \in \mathbb{C}^M, \\
        \mathbf{a}_{\mathtt{R}}(\mathbf{r},\theta) = [\mathrm{e}^{j \kappa r_1 \sin\theta}, \mathrm{e}^{j \kappa r_2 \sin\theta}, \dots, \mathrm{e}^{j \kappa r_N \sin\theta}]^\transpose \in \mathbb{C}^N,
    \end{align}
\end{subequations}
\end{small}%
where $\kappa = 2\pi/\lambda$, $\lambda$, and $\theta$ are the wavenumber, wavelength, and elevation angle, respectively. The numbers of paths are denoted by $L_{\mathtt{T}}$ and $L_{\mathtt{R}}$ for the transmit and receive MAs, respectively. We next define transmit and receive field response matrices (FRMs) as the stacks of the steering vectors:

\begin{small}
\vspace{-0.3cm}
\begin{subequations}\label{eq:frm_def}
    \begin{align}
        \mathbf{G}(\mathbf{t}) = [\mathbf{a}_{\mathtt{T}}(\mathbf{t},\theta_1^{\mathtt{T}}), \mathbf{a}_{\mathtt{T}}(\mathbf{t},\theta_2^{\mathtt{T}}), \dots, \mathbf{a}_{\mathtt{T}}(\mathbf{t},\theta_{L_{\mathtt{T}}}^{\mathtt{T}})]^\transpose \in \mathbb{C}^{L_{\mathtt{T}} \times M}, \\
        \mathbf{F}(\mathbf{r}) = [\mathbf{a}_{\mathtt{R}}(\mathbf{r},\theta_1^{\mathtt{R}}), \mathbf{a}_{\mathtt{R}}(\mathbf{r},\theta_2^{\mathtt{R}}), \dots, \mathbf{a}_{\mathtt{R}}(\mathbf{r},\theta_{L_{\mathtt{R}}}^{\mathtt{R}})]^\transpose \in \mathbb{C}^{L_{\mathtt{R}} \times N},
    \end{align}
\end{subequations}
\end{small}%
respectively. Here, $\theta_p^{\mathtt{T}}$ $(p=1,2,\dots,L_{\mathtt{T}})$ and $\theta_q^{\mathtt{R}}$ $(q=1,2,\dots,L_{\mathtt{R}})$ represent the elevation angles of departure (AoDs) and angles of arrival (AoAs), respectively. The path response matrix (PRM) is further defined as $\mathbf{\Sigma}\in\mathbb{C}^{L_{\mathtt{T}} \times L_{\mathtt{R}}}$, where $[\mathbf{\Sigma}]_{p,q}$ denotes the channel response between the $p$-th transmit and $q$-th receive paths. Therefore, the overall channel matrix between the transmitter and receiver is expressed as
\begin{equation}
    \tilde{\mathbf{H}}(\mathbf{t},\mathbf{r}) = \mathbf{F}^\hermconj(\mathbf{r}) \mathbf{\Sigma} \mathbf{G}(\mathbf{t}).
\end{equation}

In existing MA literature, MC effects are generally ignored, and corresponding signal models are given by~\cite{maMIMOCapacityCharacterization2024,zhuModelingPerformanceAnalysis2024,tangSecureMIMOCommunication2025,dingEnergyEfficiencyMaximization2025}
\begin{equation}\label{eq:signal_model}
    \mathbf{y} = \tilde{\mathbf{H}}(\mathbf{t},\mathbf{r}) \mathbf{x} + \mathbf{n},
\end{equation}
where $\mathbf{y}\in\mathbb{C}^{N}$, $\mathbf{x}\in\mathbb{C}^{M}$, and $\mathbf{n}\sim\mathcal{CN}(\mathbf{0},\sigma^2\mathbf{I}_N)$ are the received signal, transmitted signal, and additive white Gaussian noise (AWGN), respectively. The covariance matrix of the transmitted signal $\mathbf{x}$ is defined as $\mathbf{Q} \triangleq \E[\mathbf{x} \mathbf{x}^\hermconj] \in \mathbb{C}^{M\times M}$, with $\mathbf{Q} \succeq \mathbf{0}$. This model implicitly assumes that the transmitted power is defined as $P_{\mathtt{T}} = \E[\mathbf{x}^\hermconj \mathbf{x}]$. In practice, however, antenna movement alters the array's radiation pattern, causing the transmitted power to differ from the power radiated into free space. Consequently, the signal model~\eqref{eq:signal_model} may violate the principle of energy conservation.

From an electromagnetic perspective, the physically radiated power is given by the surface integral of the Poynting vector over the closed surface $\partial S$ enclosing the transmitter:~\cite{morrisSuperdirectivityMIMOSystems2005}
\begin{equation}\label{eq:radiated_power}
    P_{\mathtt{rad}} = \frac{1}{2}\oiint_{\partial S} \E[\Re\{\vec{\mathcal{E}} \times \vec{\mathcal{H}}^\star\}] \cdot \mathrm{d}\hat{\mathbf{n}},
\end{equation}
where $\vec{\mathcal{E}}$ and $\vec{\mathcal{H}}^\star$ denote the electric field and the conjugate of the magnetic field, respectively, and $\hat{\mathbf{n}}$ is the unit normal vector to the surface $\partial S$. If we omit scalar coefficients, assume isotropic antenna elements, and choose $\partial S$ as a sphere with radius $r$ centered at the origin of the transmit coordinate system, then~\eqref{eq:radiated_power} can be simplified to

\begin{small}
\vspace{-0.3cm}
\begin{align}\label{eq:radiated_power_simplified}
    P_{\mathtt{rad}} &= \E\Bigg[\mathbf{x}^\hermconj \underbrace{\left(\int_{0}^{\pi}\int_{0}^{2\pi}\mathbf{a}_{\mathtt{T}}(\mathbf{t},\theta)\mathbf{a}_{\mathtt{T}}^\hermconj(\mathbf{t},\theta) \sin\theta \mathrm{d}\phi \mathrm{d}\theta\right)}_{\mathbf{C}_{\mathtt{T}}(\mathbf{t})} \mathbf{x}\Bigg].
\end{align}
\end{small}%
The matrix $\mathbf{C}_{\mathtt{T}}(\mathbf{t})$ is referred to as the transmit \textbf{MC matrix} for isotropic antenna elements. Similarly, we define the receive MC matrix as $\mathbf{C}_{\mathtt{R}}(\mathbf{r})$. The elements of $\mathbf{C}_{\mathtt{T}}(\mathbf{t})$ and $\mathbf{C}_{\mathtt{R}}(\mathbf{r})$ are given by

\begin{small}
\vspace{-0.3cm}
\begin{subequations}\label{eq:mc_matrix_def}
    \begin{align}
        [\mathbf{C}_{\mathtt{T}}(\mathbf{t})]_{m,m^\prime} &= \mathrm{sinc}(\kappa(t_m-t_{m^\prime})), \quad m,m^\prime = 1, 2, \dots, M, \\
        [\mathbf{C}_{\mathtt{R}}(\mathbf{r})]_{n,n^\prime} &= \mathrm{sinc}(\kappa(r_n-r_{n^\prime})), \quad n,n^\prime = 1, 2, \dots, N,
    \end{align}
\end{subequations}
\end{small}%
respectively, where $\mathrm{sinc}(x) = \sin(x)/x$. Obviously, the radiated power equals the transmitted power, $P_{\mathtt{rad}} = P_{\mathtt{T}}$, if and only if $\mathbf{C}_{\mathtt{T}}(\mathbf{t}) = \mathbf{I}_M$.
\begin{remark}
    In conventional MIMO systems, antenna elements are typically spaced with integer multiples of half a wavelength. With isotropic antennas, this configuration intentionally renders the MC matrices as identity matrices, i.e., $\mathbf{C}_{\mathtt{T}} = \mathbf{I}_M$ and $\mathbf{C}_{\mathtt{R}} = \mathbf{I}_N$, under which the MC effects do not exist. In contrast, MA systems allow antenna elements to move off the half-wavelength grid, making MC inevitable. Existing studies on MA systems often ignore MC effects by enforcing a minimum antenna spacing of half a wavelength. Although the off-diagonal elements of $\mathbf{C}_{\mathtt{T}}(\mathbf{t})$ and $\mathbf{C}_{\mathtt{R}}(\mathbf{r})$ become relatively small under this constraint, MC still exists. Moreover, such simplifications overlook the influence of antenna movement on MC, leaving the impact of MC in MA systems unexplored.
\end{remark}

\subsection{Circuit-Based Communication Model}
\begin{figure}[!t]
    \centering
    \includegraphics[width=0.45\textwidth]{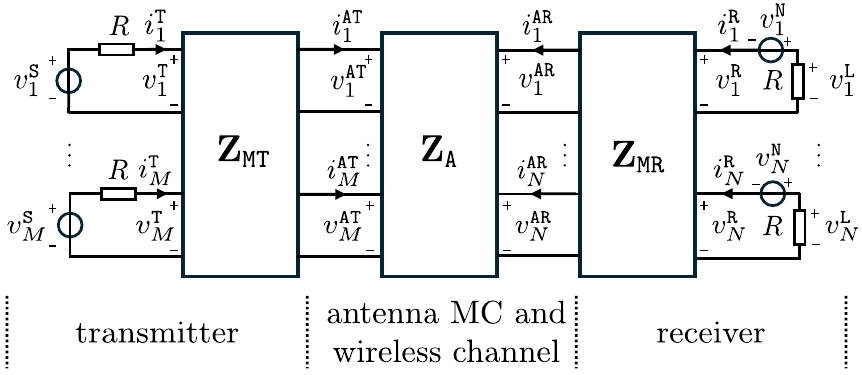}
    \caption{Multiport circuit representation of the MA system with MC.}
    \label{fig:circuit_model}
\end{figure}

As discussed above, if MC effects are ignored in MA systems, there is a mismatch between the radiated and transmitted powers. Therefore, we must adopt the circuit-based communication model~\cite{ivrlacCircuitTheoryCommunication2010}. Under the assumption of perfect transmit power matching, receive noise matching, and isotropic background radiation, the simplified model is illustrated in Fig.~\ref{fig:circuit_model}. The transmit and receive MC induced by antennas are characterized by the impedance matrix $\mathbf{Z}_{\mathtt{A}}$:
\begin{equation}\label{eq:antenna_mc_model}
    \begin{bmatrix}
        \mathbf{v}_{\mathtt{AT}} \\
        \mathbf{v}_{\mathtt{AR}}
    \end{bmatrix}
    =
    \underbrace{\begin{bmatrix}
        \mathbf{Z}_{\mathtt{AT}} & \mathbf{0}_{M\times N} \\
        \mathbf{Z}_{\mathtt{ATR}} & \mathbf{Z}_{\mathtt{AR}}
    \end{bmatrix}}_{\mathbf{Z}_{\mathtt{A}}}
    \begin{bmatrix}
        \mathbf{i}_{\mathtt{AT}} \\
        \mathbf{i}_{\mathtt{AR}}
    \end{bmatrix},
\end{equation}
where $\mathbf{i}_{\mathtt{AT}} \triangleq [i_1^{\mathtt{AT}}, i_2^{\mathtt{AT}}, \dots, i_M^{\mathtt{AT}}]^\transpose$ and $\mathbf{v}_{\mathtt{AT}} \triangleq [v_1^{\mathtt{AT}}, v_2^{\mathtt{AT}}, \dots, v_M^{\mathtt{AT}}]^\transpose$ (and $\mathbf{i}_{\mathtt{AR}} \triangleq [i_1^{\mathtt{AR}}, i_2^{\mathtt{AR}}, \dots, i_N^{\mathtt{AR}}]^\transpose$ and $\mathbf{v}_{\mathtt{AR}} \triangleq [v_1^{\mathtt{AR}}, v_2^{\mathtt{AR}}, \dots, v_N^{\mathtt{AR}}]^\transpose$) are the currents and voltages at the input of the transmit MAs (and the output of the receive MAs), respectively. The matrices $\mathbf{Z}_{\mathtt{AT}}$ and $\mathbf{Z}_{\mathtt{AR}}$ denote the transmit and receive impedance matrices, respectively, where $\Re\{\mathbf{Z}_{\mathtt{AT}}\} = \mathbf{C}_{\mathtt{T}}(\mathbf{t})$ and $\Re\{\mathbf{Z}_{\mathtt{AR}}\} = \mathbf{C}_{\mathtt{R}}(\mathbf{r})$. The transimpedance matrix $\mathbf{Z}_{\mathtt{ATR}}$ characterizes the response of the wireless channel and is equivalent to the channel matrix $\tilde{\mathbf{H}}(\mathbf{t},\mathbf{r})$, up to constant scaling factors. The contribution of the receive currents $\mathbf{i}_{\mathtt{AR}}$ to the transmit voltages $\mathbf{v}_{\mathtt{AT}}$ is ignored under the unilateral approximation~\cite{ivrlacCircuitTheoryCommunication2010}.

To ensure efficient power transfer from the signal generator $\mathbf{v}_{\mathtt{S}} \triangleq [v^{\mathtt{S}}_1, v^{\mathtt{S}}_2, \dots, v^{\mathtt{S}}_M]^\transpose$ to the MAs, a lossless impedance matching network $\mathbf{Z}_{\mathtt{MT}} \in \mathbb{C}^{2M\times 2M}$ is employed at the transmitter, as given by~\cite[Eq.~(103)]{ivrlacCircuitTheoryCommunication2010}. Similarly, a lossless impedance matching network $\mathbf{Z}_{\mathtt{MR}} \in \mathbb{C}^{2N\times 2N}$ is employed at the receiver to maximize the receive signal-to-noise ratio (SNR), with the expression provided in~\cite[Eq.~(84)]{ivrlacCircuitTheoryCommunication2010}. Using Th\'evenin's theorem~\cite{ivrlacCircuitTheoryCommunication2010} at the receiver side of Fig.~\ref{fig:circuit_model} yields the input-output relation of the system:
\begin{equation}\label{eq:system_model_mc}
    \underbrace{\tfrac{1}{2\sqrt{R}}\mathbf{v}_{\mathtt{L}}}_{\mathbf{y}}
    =
    \underbrace{\mathbf{C}_{\mathtt{R}}^{-1/2}(\mathbf{r})
    \tilde{\mathbf{H}}(\mathbf{t},\mathbf{r})
    \mathbf{C}_{\mathtt{T}}^{-1/2}(\mathbf{t})}_{\mathbf{H}(\mathbf{t},\mathbf{r})}
    \underbrace{\tfrac{1}{2\sqrt{R}}\mathbf{v}_{\mathtt{S}}}_{\mathbf{x}}
    +
    \underbrace{\tfrac{-1}{\sqrt{\Re\{Z_{\mathtt{opt}}\}}}\mathbf{v}_{\mathtt{N}}}_{\mathbf{n}},
\end{equation}
where $\mathbf{v}_{\mathtt{L}} \triangleq [v^{\mathtt{L}}_1, v^{\mathtt{L}}_2, \dots, v^{\mathtt{L}}_N]^\transpose$, $\mathbf{v}_{\mathtt{N}} \triangleq [v^{\mathtt{N}}_1, v^{\mathtt{N}}_2, \dots, v^{\mathtt{N}}_N]^\transpose$, and $\mathbf{H}(\mathbf{t},\mathbf{r})$ denote the load voltage vector, noise voltage vector, and equivalent channel matrix, respectively. $\mathbf{x}$, $\mathbf{y}$, and $\mathbf{n}$ represent the transmitted signal, received signal, and noise, respectively, consistent with~\eqref{eq:signal_model}. The radiated power $P_{\mathtt{rad}}$ now equals the transmitted power $P_{\mathtt{T}}$ with MC taken into account:
\begin{equation}
    P_{\mathtt{rad}}
    = \E[\mathbf{i}_{\mathtt{AT}}^\hermconj \mathbf{C}_{\mathtt{T}}(\mathbf{t}) \mathbf{i}_{\mathtt{AT}}]
    = \E\!\left[\tfrac{1}{4R}\mathbf{v}_{\mathtt{S}}^\hermconj \mathbf{v}_{\mathtt{S}}\right]
    = \E[\mathbf{x}^\hermconj \mathbf{x}]
    = P_{\mathtt{T}},
\end{equation}
where the second equality follows from~\cite[Eq.~(103)]{ivrlacCircuitTheoryCommunication2010}.

\subsection{Problem Formulation}
We aim to investigate the impact of MC on the theoretical capacity limit of MA-enabled MIMO systems based on the circuit-based signal model~\eqref{eq:system_model_mc}. The MIMO channel capacity is given by
\begin{equation}\label{eq:capacity}
    C(\mathbf{t},\mathbf{r}) = \max_{\substack{\mathbf{Q}: \mathbf{Q} \succeq \mathbf{0} \\ \trace(\mathbf{Q})\leq P_{\max}}} \log \det\left(\mathbf{I}_N + \tfrac{1}{\sigma^2}
        \mathbf{H}(\mathbf{t},\mathbf{r}) \mathbf{Q} \mathbf{H}^\hermconj(\mathbf{t},\mathbf{r})\right).
\end{equation}
Unlike conventional MIMO systems, the capacity $C(\mathbf{t},\mathbf{r})$ is a function of transmit and receive MA positions $\mathbf{t}$ and $\mathbf{r}$. Thus, we formulate the capacity maximization problem as
\begin{subequations}\label{eq:se_max_problem}
    \begin{align}
        \max_{\mathbf{Q},\mathbf{t},\mathbf{r}} \quad &
        \log \det\!\left(\mathbf{I}_N + \tfrac{1}{\sigma^2}
        \mathbf{H}(\mathbf{t},\mathbf{r}) \mathbf{Q} \mathbf{H}^\hermconj(\mathbf{t},\mathbf{r})\right), \label{eq:se_max_obj}\\
        \text{s.t.} \quad &
        \mathbf{Q} \succeq \mathbf{0}, \label{eq:se_max_q_psd}\\
        &
        \trace(\mathbf{Q}) \leq P_{\max}, \label{eq:se_max_power}\\
        &
        0 \leq t_m \leq D_{\mathtt{T}}, \quad m = 1,2,\dots,M, \label{eq:se_max_t_range}\\
        &
        0 \leq r_n \leq D_{\mathtt{R}}, \quad n = 1,2,\dots,N, \label{eq:se_max_r_range}\\
        &
        t_m - t_{m-1} \geq d_{\min}, \quad m = 2,3,\dots,M, \label{eq:se_max_t_spacing}\\
        &
        r_n - r_{n-1} \geq d_{\min}, \quad n = 2,3,\dots,N. \label{eq:se_max_r_spacing}
    \end{align}
\end{subequations}
Here, constraint~\eqref{eq:se_max_q_psd} enforces the positive semidefiniteness of $\mathbf{Q}$, and~\eqref{eq:se_max_power} limits the total transmit power to $P_{\max}$. Constraints~\eqref{eq:se_max_t_range} and~\eqref{eq:se_max_r_range} restrict the transmit and receive MAs within their respective movement ranges. Constraints~\eqref{eq:se_max_t_spacing} and~\eqref{eq:se_max_r_spacing} ensure a minimum spacing between adjacent MAs.\footnote{In MA systems without MC~\cite{maMIMOCapacityCharacterization2024,zhuModelingPerformanceAnalysis2024}, the spacing constraints~\eqref{eq:se_max_t_spacing} and~\eqref{eq:se_max_r_spacing} are typically imposed to mitigate MC among antenna elements, where the minimum spacing $d_{\min}$ is usually set exactly as $\lambda/2$. Although MC is explicitly considered in this paper, these constraints remain necessary to prevent physical collisions between MAs with much smaller $d_{\min}$ than $\lambda/2$.}

The optimization problem~\eqref{eq:se_max_problem} is challenging to solve due to the non-concavity of its objective function. Compared with the capacity maximization problem that ignores MC~\cite{maMIMOCapacityCharacterization2024}, the presence of the additional MC matrices $\mathbf{C}_{\mathtt{T}}(\mathbf{t})$ and $\mathbf{C}_{\mathtt{R}}(\mathbf{r})$ makes the problem more challenging. In particular, the inverse square roots $\mathbf{C}^{-1/2}_{\mathtt{T}}(\mathbf{t})$ and $\mathbf{C}^{-1/2}_{\mathtt{R}}(\mathbf{r})$ lack closed form expressions, making gradient derivations analytically intractable.

\begin{remark}\label{rem:capacity_gain}
    The potential capacity gains of leveraging MC in MA systems are twofold. First, in MA systems, MC-induced superdirectivity can result in higher effective transmit and receive power~\cite{morrisSuperdirectivityMIMOSystems2005}. When MAs move off the $\lambda/2$ grid, the eigenvalues of the MC matrices $\mathbf{C}_{\mathtt{T}}(\mathbf{t})$ and $\mathbf{C}_{\mathtt{R}}(\mathbf{r})$ are no longer uniformly equal to $1$. Smaller eigenvalues of these matrices yield larger eigenvalues of $\mathbf{C}_{\mathtt{T}}^{-1/2}(\mathbf{t})$ and $\mathbf{C}_{\mathtt{R}}^{-1/2}(\mathbf{r})$, which in turn leads to higher transmit and receive power density along the directions of $\{\theta_p^{\mathtt{T}}\}_{p=1}^{L_{\mathtt{T}}}$ and $\{\theta_q^{\mathtt{R}}\}_{q=1}^{L_{\mathtt{R}}}$, respectively. Second, unlike conventional MIMO systems with MC, MA makes the MC matrices $\mathbf{C}_{\mathtt{T}}$ and $\mathbf{C}_{\mathtt{R}}$ designable. Position optimization shapes the MC matrices to align with the eigenmodes of the wireless channel, resulting in a more uniform eigenvalue spectrum of the SNR matrix $\mathbf{H}\mathbf{Q}\mathbf{H}^\hermconj/\sigma^2$ and higher capacity. Therefore, solving problem~\eqref{eq:se_max_problem} is the key to fully exploiting MC in MA systems.
\end{remark}

\section{BCA-Based Algorithm}
In this section, we propose a BCA-based algorithm to iteratively solve problem~\eqref{eq:se_max_problem}. In each iteration, each set of parameters is optimized in an alternating manner with others fixed. Specifically, the transmit covariance matrix $\mathbf{Q}$ is optimized using the water-filling algorithm~\cite{tse2005fundamentals}, and the MA positions $\mathbf{t}$ and $\mathbf{r}$ are optimized using the TRM~\cite{yuan2015recent}.

\subsection{Update of $\mathbf{Q}$}\label{subsec:q}
In this step, we optimize the transmit covariance matrix $\mathbf{Q}$ while keeping the MA positions $\mathbf{t}$ and $\mathbf{r}$ fixed. This subproblem is concave and can be efficiently solved using the water-filling algorithm~\cite{tse2005fundamentals}. Let the singular value decomposition (SVD) of $\mathbf{H}(\overline{\mathbf{t}},\overline{\mathbf{r}})$ be expressed as $\mathbf{H}(\overline{\mathbf{t}},\overline{\mathbf{r}}) = \overline{\mathbf{U}} \overline{\mathbf{\Lambda}} \overline{\mathbf{V}}^\hermconj$, where $\overline{\mathbf{\Lambda}} = \mathrm{diag}(\lambda_1, \lambda_2, \dots, \lambda_S) \in \mathbb{C}^{S\times S}$, $\overline{\mathbf{U}} \in \mathbb{C}^{N\times S}$, and $\overline{\mathbf{V}} \in \mathbb{C}^{M\times S}$ denote the singular values, left singular vectors, and right singular vectors of $\mathbf{H}(\overline{\mathbf{t}},\overline{\mathbf{r}})$, respectively. Here, $\overline{\mathbf{t}}$ and $\overline{\mathbf{r}}$ represent the temporarily optimized transmit and receive MA positions from the previous iteration, and $S$ denotes the rank of $\mathbf{H}(\overline{\mathbf{t}},\overline{\mathbf{r}})$.

According to the water-filling principle, the transmit power allocated to the $s$-th eigenchannel with singular value $\lambda_s$ is
\begin{equation}
    P_s = \max\left(0,\, \mu - \tfrac{\sigma^2}{\lambda_s^2}\right), \qquad s = 1,2,\dots,S,
\end{equation}
where $\mu$ is the water level chosen to satisfy $\sum_{s=1}^{S} P_s = P_{\max}$. Thus, the optimal transmit covariance matrix is given by
\begin{equation}\label{eq:q_opt}
    \mathbf{Q} = \overline{\mathbf{V}}\, \mathrm{diag}(P_1, P_2, \dots, P_S)\, \overline{\mathbf{V}}^\hermconj.
\end{equation}

\subsection{Update of $t_m$}\label{subsec:t_m}
In this step, we optimize the $m$-th transmit MA position $t_m$ while keeping the transmit covariance matrix $\mathbf{Q}$, the receive MA positions $\mathbf{r}$, and the other transmit MA positions $t_{m^\prime}\ (m^\prime \neq m)$ fixed. The optimization problem~\eqref{eq:se_max_problem} can be reformulated as
\begin{equation}\label{eq:t_m_problem}
    \max_{t_m}\ h_\mathtt{T}(t_m),\quad \text{s.t.}\ \overline{t}_{m-1}+d_{\min}\leq t_m \leq \overline{t}_{m+1}-d_{\min},
\end{equation}
where we define $\overline{t}_0 = -d_{\min}$ and $\overline{t}_{M+1} = D_{\mathtt{T}}+d_{\min}$ for convenience. The objective function $h_\mathtt{T}(t_m)$ equals $C(\mathbf{t},\mathbf{r})$ in~\eqref{eq:capacity} with only $t_m$ being the variable, and it remains non-concave. Determining update step sizes that ensure monotonic increases in $h_\mathtt{T}(t_m)$ is challenging, as its curvature upper bound cannot be explicitly derived. Therefore, we adopt the TRM~\cite{yuan2015recent} to solve problem~\eqref{eq:t_m_problem}, as it guarantees convergence using only the first- and second-order partial derivatives of $h_\mathtt{T}(t_m)$ and remains effective even with indefinite curvature.

\begin{figure}[tb]
    \centering
    \includegraphics[width=0.4\textwidth]{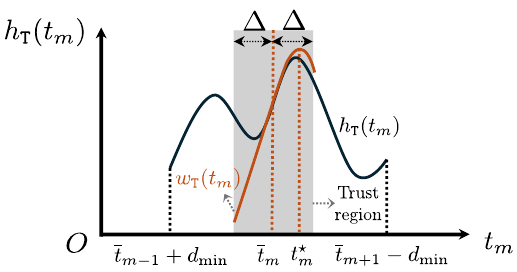}
    \caption{TRM for updating $t_m$.}
    \label{fig:trust_region}
\end{figure}

Fig.~\ref{fig:trust_region} illustrates the TRM for updating $t_m$. The trust region (TR) is defined as the set of candidate values centered at $\overline{t}_m$ with a radius of $\Delta$. Within the intersection of the TR and the feasible set $\mathcal{T}_m$ defined by the constraint of~\eqref{eq:t_m_problem}, the objective function $h_\mathtt{T}(t_m)$ is locally approximated by a quadratic function $w_{\mathtt{T}}(t_m)$:

\begin{small}
\vspace{-0.3cm}
\begin{equation}\label{eq:w_t_m}
    w_{\mathtt{T}}(t_m)
    = \tfrac{1}{2} h^{\prime\prime}_\mathtt{T}(\overline{t}_m)(t_m - \overline{t}_m)^2
    + h^{\prime}_\mathtt{T}(\overline{t}_m)(t_m - \overline{t}_m)
    + h_\mathtt{T}(\overline{t}_m),
\end{equation}
\end{small}%
where $h_\mathtt{T}^{\prime}$ and $h_\mathtt{T}^{\prime\prime}$ denote the first- and second-order derivatives of $h_\mathtt{T}(t_m)$, respectively. The computation of these derivatives will be detailed later. The feasible set $\mathcal{T}_m$ is defined as
\begin{equation}
    \mathcal{T}_m \triangleq 
    [\max(\overline{t}_m - \Delta,\overline{t}_{m-1} + d_{\min}),
     \min(\overline{t}_m + \Delta,\overline{t}_{m+1} - d_{\min})].
\end{equation}
Since $w_{\mathtt{T}}(t_m)$ is quadratic, its optimal solution is given by
\begin{equation}\label{eq:t_m_star}
    t_m^\star = \mathrm{Proj}_{\mathcal{T}_m}\!\left(
    \overline{t}_m - 
    h^{\prime}_\mathtt{T}(\overline{t}_m)/h^{\prime\prime}_\mathtt{T}(\overline{t}_m)
    \right),
\end{equation}
where $\mathrm{Proj}_{\mathcal{T}_m}(\cdot)$ denotes the projection operator onto the set $\mathcal{T}_m$. If $w_{\mathtt{T}}(t_m)$ provides a good local approximation of $h_\mathtt{T}(t_m)$, the optimum $t_m^\star$ of $w_{\mathtt{T}}(t_m)$ leads to an improvement in $h_\mathtt{T}(t_m)$. The approximation quality is evaluated by the ratio
\begin{equation}\label{eq:rho}
    \rho = 
    \frac{h_\mathtt{T}(\overline{t}_m) - h_\mathtt{T}(t_m^\star)}
         {w_\mathtt{T}(\overline{t}_m) - w_{\mathtt{T}}(t_m^\star)}.
\end{equation}
We define two constant thresholds, $\rho_1$ and $\rho_2$, satisfying $0 < \rho_1 < \rho_2 < 1$, to evaluate the approximation quality. If $\rho > \rho_2$, $w_{\mathtt{T}}(t_m)$ is considered a good local approximation of $h_\mathtt{T}(t_m)$. In this case, the update is accepted by setting $t_m = t_m^\star$. Moreover, if $t_m^\star$ lies on the boundary of the TR, the radius $\Delta$ increases to $\nu_1\Delta$ for the next iteration. If $\rho_1 < \rho \leq \rho_2$, the update is accepted while keeping $\Delta$ unchanged. If $\rho \leq \rho_1$, the update is rejected, and the radius is reduced to $\Delta/\nu_2$.

The key to applying the TRM lies in computing the first- and second-order derivatives of $h_\mathtt{T}(t_m)$. Using the matrix chain rule~\cite{petersen2008matrix}, we obtain the first- and second-order derivatives of $h_\mathtt{T}(t_m)$ as
\begin{align}
    h^{\prime}_\mathtt{T}(t_m) &= \tfrac2{\sigma^2}\Re\left\{\trace\left[\mathbf{\Phi} \tfrac{\partial \mathbf{H}}{\partial t_m} \overline{\mathbf{Q}}\mathbf{H}^\hermconj\right] \right\},\label{eq:h_prime_t_m}\\
    h^{\prime\prime}_\mathtt{T}(t_m) &= \tfrac2{\sigma^2}\Re\left\{\trace\left[\mathbf{\Phi}\left(\tfrac{\partial^2 \mathbf{H}}{\partial t_m^2} \overline{\mathbf{Q}}\mathbf{H}^\hermconj + \tfrac{\partial \mathbf{H}}{\partial t_m} \overline{\mathbf{Q}} \tfrac{\partial \mathbf{H}^\hermconj}{\partial t_m}\right)\right. \right.\nonumber \\
    &\hspace{3em}-\sigma^{-2}\mathbf{\Phi}\tfrac{\partial \mathbf{H}}{\partial t_m} \overline{\mathbf{Q}}\mathbf{H}^\hermconj \mathbf{\Phi}\tfrac{\partial \mathbf{H}}{\partial t_m}\overline{\mathbf{Q}}\mathbf{H}^\hermconj \nonumber \\
    &\hspace{3em}-\left.\left.\sigma^{-2} \mathbf{\Phi}\tfrac{\partial \mathbf{H}}{\partial t_m} \overline{\mathbf{Q}}\mathbf{H}^\hermconj \mathbf{\Phi}\mathbf{H}\overline{\mathbf{Q}}\tfrac{\partial \mathbf{H}^\hermconj}{\partial t_m}\right]\right\},\label{eq:h_prime2_t_m}
\end{align}
where $\mathbf{\Phi} \triangleq \left(\mathbf{I}_N + \sigma^{-2}\mathbf{H}\overline{\mathbf{Q}}\mathbf{H}^\hermconj\right)^{-1}$ and $\overline{\mathbf{Q}}$ denotes the temporarily optimized $\mathbf{Q}$. The first- and second-order derivatives of $\mathbf{H}$ with respect to (w.r.t.) $t_m$ are given by

\begin{small}
\vspace{-0.3cm}
\begin{align}
    \frac{\partial \mathbf{H}}{\partial t_m} &= \overline{\mathbf{C}}_{\mathtt{R}}^{-\frac12} \overline{\mathbf{F}}^\hermconj \mathbf{\Sigma} \Big[\frac{\partial \mathbf{G}}{\partial t_m} \mathbf{C}_{\mathtt{T}}^{-\frac12} + \mathbf{G} \frac{\partial \mathbf{C}_{\mathtt{T}}^{-1/2}}{\partial t_m}\Big], \label{eq:partial_h_t_m}\\
    \frac{\partial^2 \mathbf{H}}{\partial t_m^2} &= \overline{\mathbf{C}}_{\mathtt{R}}^{-\frac12} \overline{\mathbf{F}}^\hermconj \mathbf{\Sigma}\Big[\frac{\partial^2 \mathbf{G}}{\partial t_m^2} \mathbf{C}_{\mathtt{T}}^{-\frac12} + 2\frac{\partial \mathbf{G}}{\partial t_m} \frac{\partial \mathbf{C}_{\mathtt{T}}^{-1/2}}{\partial t_m} + \mathbf{G} \frac{\partial^2 \mathbf{C}_{\mathtt{T}}^{-1/2}}{\partial t_m^2}\Big], \label{eq:partial2_h_t_m}
\end{align}
\end{small}%
respectively. Here $\overline{\mathbf{C}}_{\mathtt{R}}$ and $\overline{\mathbf{F}}$ are taken from the previous iteration. In~\eqref{eq:partial_h_t_m} and~\eqref{eq:partial2_h_t_m}, the elements of the first- and second-order derivatives of $\mathbf{G}$ w.r.t. $t_m$ are
\begin{align}
    \Big[\frac{\partial \mathbf{G}}{\partial t_m}\Big]_{p, m^{\prime}} &= \left\{
        \begin{aligned}
            &j\kappa\sin\theta_p \mathrm{e}^{j\kappa t_{m^\prime}}, &  m^\prime = m,\\
            &0, & m^\prime \neq m,
        \end{aligned}\right.\label{eq:partial_g_t_m}\\
    \Big[\frac{\partial^2 \mathbf{G}}{\partial t_m^2}\Big]_{p, m^{\prime}} &= \left\{
        \begin{aligned}
            &-\kappa^2\sin^2\theta_p \mathrm{e}^{j\kappa t_{m^\prime}}, &  m^\prime = m,\\
            &0, & m^\prime \neq m,
        \end{aligned}\right.\label{eq:partial2_g_t_m}
\end{align}
respectively. Note that in~\eqref{eq:partial_h_t_m} and~\eqref{eq:partial2_h_t_m}, the first- and second-order partial derivatives of $\mathbf{C}_{\mathtt{T}}^{-1/2}$ w.r.t. $t_m$ are analytically intractable. Instead, we first construct equations characterizing the partial derivatives of $\mathbf{C}_{\mathtt{T}}^{-1/2}$ and then obtain their exact values by solving these equations. We first derive the equation characterizing $\frac{\partial \mathbf{C}_{\mathtt{T}}^{-1/2}}{\partial t_m}$. By taking the partial derivative on both sides of the identity $\mathbf{C}_{\mathtt{T}}^{-1/2} \mathbf{C}_{\mathtt{T}} \mathbf{C}_{\mathtt{T}}^{-1/2} = \mathbf{I}_M$ w.r.t. $t_m$, we have
\begin{equation}\label{eq:sylvester_equation_1}
    \frac{\partial \mathbf{C}_{\mathtt{T}}^{-1/2}}{\partial t_m} \mathbf{C}_{\mathtt{T}}^{\frac12}
    + \mathbf{C}_{\mathtt{T}}^{\frac12} \frac{\partial \mathbf{C}_{\mathtt{T}}^{-1/2}}{\partial t_m}
    = -\mathbf{C}_{\mathtt{T}}^{-\frac12} \frac{\partial \mathbf{C}_{\mathtt{T}}}{\partial t_m} \mathbf{C}_{\mathtt{T}}^{-\frac12}.
\end{equation}
The elements of $\frac{\partial \mathbf{C}_{\mathtt{T}}}{\partial t_m}$ are given by
\begin{small}
\begin{align}\label{eq:partial_c_t}
    &\Big[\frac{\partial \mathbf{C}_{\mathtt{T}}}{\partial t_m}\Big]_{m_1, m_2} \nonumber \\
    =&
    \begin{cases}
        \dfrac{\cos\kappa(t_{m_1}-t_{m_2})}{t_{m_1}-t_{m_2}}
        - \dfrac{\sin\kappa(t_{m_1}-t_{m_2})}{\kappa(t_{m_1}-t_{m_2})^2}, & m_1 \neq m_2 = m,\\
        \Big[\dfrac{\partial \mathbf{C}_{\mathtt{T}}}{\partial t_m}\Big]_{m_2, m_1}, & m_2 \neq m_1 = m,\\
        0, & \text{otherwise}.
    \end{cases}
\end{align}
\end{small}%
Similarly, the equation characterizing $\frac{\partial^2 \mathbf{C}_{\mathtt{T}}^{-1/2}}{\partial t_m^2}$ can be obtained by taking the partial derivative on both sides of~\eqref{eq:sylvester_equation_1} w.r.t. $t_m$:
\begin{align}\label{eq:sylvester_equation_2}
    \frac{\partial^2 \mathbf{C}_{\mathtt{T}}^{-1/2}}{\partial t_m^2} \mathbf{C}_{\mathtt{T}}^{-\frac12}
    + \mathbf{C}_{\mathtt{T}}^{-\frac12} \frac{\partial^2 \mathbf{C}_{\mathtt{T}}^{-1/2}}{\partial t_m^2}
    = &-\mathbf{C}_{\mathtt{T}}^{-\frac12} \frac{\partial^2 \mathbf{C}_{\mathtt{T}}}{\partial t_m^2} \mathbf{C}_{\mathtt{T}}^{-\frac12} \\ \nonumber
    - \mathbf{C}_{\mathtt{T}}^{-\frac12} \frac{\partial \mathbf{C}_{\mathtt{T}}}{\partial t_m} \frac{\partial \mathbf{C}_{\mathtt{T}}^{-1/2}}{\partial t_m}
    &- \frac{\partial \mathbf{C}_{\mathtt{T}}^{-1/2}}{\partial t_m} \frac{\partial \mathbf{C}_{\mathtt{T}}}{\partial t_m} \mathbf{C}_{\mathtt{T}}^{-\frac12} \\ \nonumber
    - \frac{\partial \mathbf{C}_{\mathtt{T}}^{1/2}}{\partial t_m} \frac{\partial \mathbf{C}_{\mathtt{T}}^{-1/2}}{\partial t_m}
    &- \frac{\partial \mathbf{C}_{\mathtt{T}}^{-1/2}}{\partial t_m} \frac{\partial \mathbf{C}_{\mathtt{T}}^{1/2}}{\partial t_m},
\end{align}
where the elements of $\frac{\partial^2 \mathbf{C}_{\mathtt{T}}}{\partial t_m^2}$ are given by
\begin{small} 
\begin{align} \label{eq:partial2_c_t}
    &\Big[\frac{\partial^2 \mathbf{C}_{\mathtt{T}}}{\partial t_m^2}\Big]_{m_1, m_2} \nonumber \\ 
    =& \begin{cases} 
        -\kappa\frac{\sin\kappa(t_{m_1}-t_{m_2})}{t_{m_1}-t_{m_2}} - 2\frac{\cos\kappa(t_{m_1}-t_{m_2})}{\kappa(t_{m_1}-t_{m_2})^2} 
        &+ 2\frac{\sin\kappa(t_{m_1}-t_{m_2})}{\kappa(t_{m_1}-t_{m_2})^3}, \\ 
        & m_1 \neq m_2 = m,\\ \Big[\dfrac{\partial^2 \mathbf{C}_{\mathtt{T}}}{\partial t_m^2}\Big]_{m_2, m_1}, & m_2 \neq m_1 = m,\\ 
        0, & \text{otherwise}.
    \end{cases} 
\end{align} 
\end{small}%
The term $\frac{\partial \mathbf{C}_{\mathtt{T}}^{1/2}}{\partial t_m}$ in~\eqref{eq:sylvester_equation_2} is characterized by the equation:
\begin{equation}\label{eq:sylvester_equation_12}
    \frac{\partial \mathbf{C}_{\mathtt{T}}^{1/2}}{\partial t_m} \mathbf{C}_{\mathtt{T}}^{\frac12}
    + \mathbf{C}_{\mathtt{T}}^{\frac12} \frac{\partial \mathbf{C}_{\mathtt{T}}^{1/2}}{\partial t_m}
    = \frac{\partial\mathbf{C}_{\mathtt{T}}}{\partial t_m}.
\end{equation}
Fortunately, Eqs.~\eqref{eq:sylvester_equation_1},~\eqref{eq:sylvester_equation_2}, and~\eqref{eq:sylvester_equation_12} all take the standard form of the Sylvester equation~\cite{higham2008functions}, which can be efficiently solved using standard solvers, e.g., the MATLAB function \texttt{sylvester}. In conclusion, the first- and second-order partial derivatives of the objective function, $h_{\mathtt{T}}^{\prime}(t_m)$ and $h_{\mathtt{T}}^{\prime\prime}(t_m)$, are obtained by first substituting the solutions of~\eqref{eq:sylvester_equation_1},~\eqref{eq:sylvester_equation_2}, and~\eqref{eq:sylvester_equation_12} into~\eqref{eq:partial_h_t_m} and~\eqref{eq:partial2_h_t_m}, and then substituting the results into~\eqref{eq:h_prime_t_m} and~\eqref{eq:h_prime2_t_m}. With these derivatives, the local approximation $w_{\mathtt{T}}(t_m)$ is constructed according to~\eqref{eq:w_t_m}. Therefore, subproblem~\eqref{eq:t_m_problem} can be readily solved using the TRM.

\begin{remark}
Although only isotropic antenna elements are considered in this paper, the proposed TRM-based algorithm can be readily extended to more practical antenna types, such as dipoles. Changing the antenna model only affects the coupling matrix $\mathbf{C}_{\mathtt{T}}$. One simply needs to rederive~\eqref{eq:partial_c_t} and~\eqref{eq:partial2_c_t}, which involve only \textbf{element-wise} derivatives.
\end{remark}

\subsection{Update of $r_n$}\label{subsec:r_n}
In this step, we aim to optimize the $n$-th receive MA position $r_n$ while keeping the transmit covariance matrix $\mathbf{Q}$, the transmit MA positions $\mathbf{t}$, and the other receive MA positions $r_{n^\prime}\ (n^\prime \neq n)$ fixed. The optimization problem~\eqref{eq:se_max_problem} can be reformulated as
\begin{equation}
    \max_{r_n}\ h_\mathtt{R}(r_n),\quad \text{s.t.}\ \overline{r}_{n-1}+d_{\min}\leq r_n \leq \overline{r}_{n+1}-d_{\min},
\end{equation}
where we define $\overline{r}_0 = -d_{\min}$ and $\overline{r}_{N+1} = D_{\mathtt{R}}+d_{\min}$. The objective function $h_{\mathtt{R}}(r_n)$ is defined as
\begin{equation}
    h_{\mathtt{R}}(r_n) = \log \det\left(\mathbf{I}_M + \tfrac{1}{\sigma^2}
    \mathbf{H}^\hermconj(r_n) \overline{\mathbf{S}} \mathbf{H}(r_n)\right).
\end{equation}
Here, $\overline{\mathbf{S}} = \overline{\mathbf{U}}\mathrm{diag}(P_1, P_2, \dots, P_S)\overline{\mathbf{U}}^\hermconj$.  
The channel reciprocity property in MIMO systems~\cite[Eq.~(21)]{maMIMOCapacityCharacterization2024} ensures that the objective $h_{\mathtt{R}}(r_n)$ equals the capacity $C$ after each update of $\mathbf{Q}$. The update of $r_n$ can be obtained using a similar TRM-based procedure as detailed in Section~\ref{subsec:t_m}. The only differences lie in the calculation of partial derivatives. The first- and second-order derivatives of $h_\mathtt{R}(r_n)$ can be obtained by substituting $\mathbf{H}$ with $\mathbf{H}^\hermconj$ and $\overline{\mathbf{Q}}$ with $\overline{\mathbf{S}}$ in~\eqref{eq:h_prime_t_m} and~\eqref{eq:h_prime2_t_m}. The first- and second-order derivatives of $\mathbf{H}$ w.r.t. $r_n$ are:

\addtolength{\topmargin}{0.05in}

\begin{small}
\vspace{-0.3cm}
\begin{equation}
    \frac{\partial \mathbf{H}}{\partial r_n} = \Big[\frac{\partial \mathbf{C}_{\mathtt{R}}^{-1/2}}{\partial r_n} \mathbf{F}^\hermconj + \mathbf{C}_{\mathtt{R}}^{-1/2} \frac{\partial \mathbf{F}^\hermconj}{\partial r_n}\Big] \mathbf{\Sigma} \overline{\mathbf{G}}\overline{\mathbf{C}}_{\mathtt{T}}^{-1/2}, \label{eq:partial_h_r_n}\\
\end{equation}
\begin{equation}
	 \frac{\partial^2 \mathbf{H}}{\partial r_n^2} = \Big[\frac{\partial^2 \mathbf{C}_{\mathtt{R}}^{-1/2}}{\partial r_n^2} \mathbf{F}^\hermconj + 2\frac{\partial \mathbf{C}_{\mathtt{R}}^{-1/2}}{\partial r_n} \frac{\partial \mathbf{F}^\hermconj}{\partial r_n} + \mathbf{C}_{\mathtt{R}}^{-\frac12} \frac{\partial^2 \mathbf{F}^\hermconj}{\partial r_n^2}\Big] \mathbf{\Sigma} \overline{\mathbf{G}}\overline{\mathbf{C}}_{\mathtt{T}}^{-\frac12}, \label{eq:partial2_h_r_n}
\end{equation}
\end{small}%
respectively. The remaining derivations follow those in~\eqref{eq:partial_g_t_m}--\eqref{eq:sylvester_equation_12}, by replacing $t_m$ with $r_n$, $\theta_p$ with $\theta_q$, $\mathbf{G}$ with $\mathbf{F}$, and the subscript (or superscript) $\mathtt{T}$ with $\mathtt{R}$.  
The overall procedure for solving the optimization problem~\eqref{eq:se_max_problem} is summarized in Algorithm~\ref{alg:bca}.

\begin{algorithm}
    \caption{BCA-based algorithm for solving problem~\eqref{eq:se_max_problem}.}\label{alg:bca}
    \begin{algorithmic}[1]
        \Require $\kappa$, $P_{\max}$, $\mathbf{\Sigma}$, $L_\mathtt{T}$, $L_\mathtt{R}$, $\{\theta_p^{\mathtt{T}}\}_{p=1}^{L_\mathtt{T}}$, $\{\theta_q^{\mathtt{R}}\}_{q=1}^{L_\mathtt{R}}$, $D_\mathtt{T}$, $D_\mathtt{R}$, $d_{\min}$.
        \State Initialize $\mathbf{t}$ and $\mathbf{r}$ with feasible values.
        \Repeat
            \State Update $\mathbf{Q}$ based on~\eqref{eq:q_opt}.
            \State Update each $t_m$ using the TRM in Section~\ref{subsec:t_m}.
            \State Update each $r_n$ using the TRM in Section~\ref{subsec:r_n}.
        \Until{the objective value $C$ converges.}
        \Ensure $\mathbf{Q}$, $\mathbf{t}$, $\mathbf{r}$.
    \end{algorithmic}
\end{algorithm}

\begin{remark}
    The convergence of Algorithm~\ref{alg:bca} is guaranteed by the non-decreasing capacity $C$ after each update of $\mathbf{Q}$, $t_m$, and $r_n$. The convergence of the proposed algorithm will be further validated in Section~\ref{sec:sim}.
\end{remark}

\section{Simulation Results}\label{sec:sim}
\begin{figure*}[tb]
    \centering
    \begin{minipage}{0.3\linewidth}
        \centering
        \includegraphics[width=\linewidth]{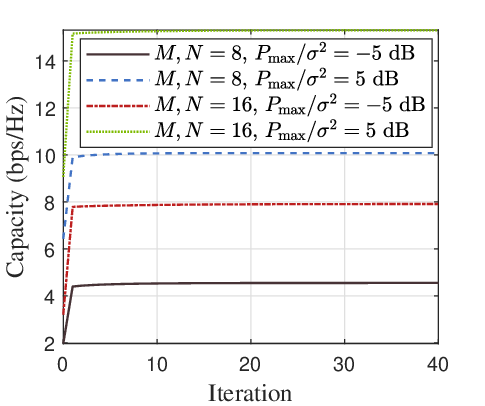}
        \caption{Convergence behaviors of Algorithm~\ref{alg:bca}.}
        \label{fig:conv}
    \end{minipage}
    \begin{minipage}{0.3\linewidth}
        \centering
        \includegraphics[width=\linewidth]{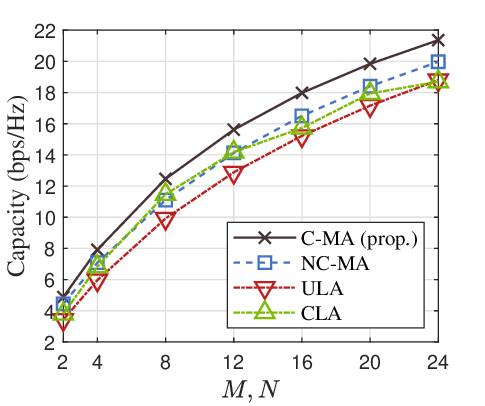}
        \caption{Impact of number of MAs ($M=N$).}
        \label{fig:ntx}
    \end{minipage}
    \begin{minipage}{0.3\linewidth}
        \centering
        \includegraphics[width=\linewidth]{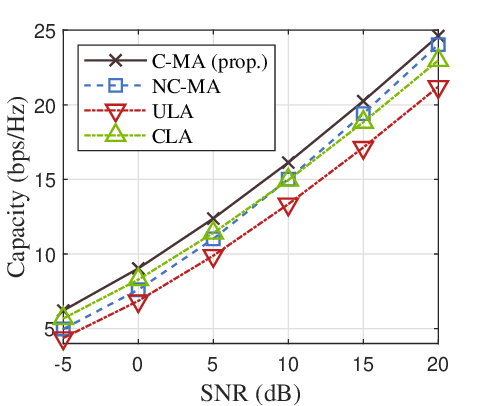}
        \caption{Impact of SNR.}
        \label{fig:snr}
    \end{minipage}
    \vspace{-0.5cm}
\end{figure*}
In this section, we evaluate the capacity improvements achieved by incorporating MC using Algorithm~\ref{alg:bca}. Unless otherwise specified, the simulation parameters are set as follows: the numbers of transmit and receive MAs are $M = N = 8$, the carrier frequency is $f_c = 28$~GHz, and the minimum antenna separation is $d_{\min} = 0.1\lambda$. The movement ranges of the transmit and receive MAs are $D_\mathtt{T} = 2M\lambda$ and $D_\mathtt{R} = 2N\lambda$, respectively. The numbers of transmit and receive paths are $L_\mathtt{T} = L_\mathtt{R} = 3$. For each transmit and receive path, the AoD and AoA are uniformly distributed over $[0, \pi)$. The SNR is set to $P_{\max}/\sigma^2 = 5$~dB. The parameters for the TRM are $\rho_1 = 0.25$, $\rho_2 = 0.75$, $\nu_1 = 2$, and $\nu_2 = 4$. All results are averaged over $1000$ independent channel realizations. We compare the capacity of the proposed algorithm, denoted by C-MA, with several baseline schemes.
\begin{itemize}
    \item \textbf{Uniform linear array (ULA)}: Fixed-position antennas (FPAs) with antenna spacing of $\lambda/2$. In this case, the MC matrix equals the identity matrix, and MC does not exist.
    \item \textbf{Compact linear array (CLA)}: FPAs with antenna spacing of $0.1\lambda$. Since the antenna spacing equals $d_{\min}$, its MC is comparable to, if not stronger than, that of C-MA.
    \item \textbf{No MC-MA (NC-MA)}~\cite{maMIMOCapacityCharacterization2024}: No MC is considered in the optimization algorithm of MA systems. The minimum antenna separation is set to $d_{\min} = \lambda/2$ to eliminate MC.
\end{itemize}

First, we evaluate the convergence behavior of Algorithm~\ref{alg:bca}, as shown in Fig.~\ref{fig:conv}. Under all setups, the capacity of C-MA increases monotonically with iterations and converges within $20$ iterations. We then evaluate the capacity gains of C-MA under different numbers of transmit and receive MAs, as shown in Fig.~\ref{fig:ntx}. C-MA consistently outperforms all baseline schemes across various antenna configurations. When $M = N = 8$, C-MA achieves approximately $12\%$ higher capacity than NC-MA. This gain arises from superdirectivity, which enables beam gains exceeding the number of antennas~\cite{ivrlacCircuitTheoryCommunication2010}, which is the theoretical limit when MC is neglected. Although CLA also exhibits superdirectivity, its capacity remains lower than that of C-MA, particularly when the antenna count is large. The movement of antennas in C-MA allows the MC matrices to be adaptively shaped to better align with the wireless channel, thereby achieving higher capacity. Finally, Fig.~\ref{fig:snr} presents the impact of SNR on capacity. C-MA achieves notable capacity improvements over all baselines, especially at low SNR, where the concavity of the logarithmic function amplifies the effect of power gains. When SNR equals $-5$~dB, the capacity gain over NC-MA is up to $25\%$.

As suggested by Remark~\ref{rem:capacity_gain}, the capacity gains of C-MA over NC-MA arise from superdirectivity effects, and now we verify it numerically using the parameters detailed in the first paragraph of this section. We only explore the transmit superdirective beam gains since the transmit and receive arrays are symmetric. The sum of radiated power density of the transmitter along the directions of $\{\theta_p^{\mathtt{T}}\}_{p=1}^{L_\mathtt{T}}$ is given by
\begin{equation}
    P_{\mathtt{trans}} = \trace\left(\mathbf{G}(\mathbf{t}) \mathbf{C}_{\mathtt{T}}^{-1/2}(\mathbf{t}) \mathbf{Q} \mathbf{C}_{\mathtt{T}}^{-1/2}(\mathbf{t}) \mathbf{G}^\hermconj(\mathbf{t})\right).
\end{equation}
In NC-MA, $P_{\mathtt{trans}} = 9.7$~W/sr, whereas in C-MA, $P_{\mathtt{trans}} = 18.4$~W/sr. Since $P_{\mathtt{trans}}$ is significantly higher in C-MA than in NC-MA, it results in higher capacity. Apart from superdirectivity, antenna movement aligns $\mathbf{C}_{\mathtt{T}}$ and $\mathbf{C}_{\mathtt{R}}$ with the wireless channel. Although the total transmitted power is higher in CLA than in C-MA due to stronger MC, its capacity is lower than that of C-MA. Next, we compute the eigenvalues of the SNR matrix $\mathbf{\Gamma} \triangleq \mathbf{H}\mathbf{Q}\mathbf{H}^\hermconj / \sigma^2$. In C-MA, the non-zero eigenvalues are $\mathbf{\lambda}(\mathbf{\Gamma}_{\text{C-MA}}) = [275.1, 29.4, 2.1]$, whereas in CLA, they are $\mathbf{\lambda}(\mathbf{\Gamma}_{\text{CLA}}) = [299.1, 21.8, 1.1]$. Although the transferred power is higher in CLA than in C-MA, i.e., $\trace(\mathbf{\Gamma}_{\text{CLA}}) > \trace(\mathbf{\Gamma}_{\text{C-MA}})$, the small eigenvalues of $\mathbf{\Gamma}_{\text{C-MA}}$ are larger than those of $\mathbf{\Gamma}_{\text{CLA}}$. Owing to the concavity of the logarithmic function, the capacity achieved by C-MA is higher than that of CLA.

\section{Conclusion and Future Work}
In this paper, we proposed leveraging MC in an MA-enabled point-to-point MIMO system. For this purpose, the system is characterized using a circuit-based communication model. We formulated the capacity maximization problem as a non-concave optimization problem and solved it using a BCA-based algorithm. For the subproblem of MA position optimization, we developed a TRM-based algorithm and derived the first- and second-order derivatives using Sylvester equations. Simulation results demonstrated that MC can be effectively exploited through superdirectivity and antenna movement. Future work includes extending the algorithm to more practical antenna types, incorporating imperfect impedance matching, and exploring multiuser scenarios.

\bibliographystyle{IEEEtran}
\begin{spacing}{0.92}
\bibliography{IEEEabrv,reference}
\end{spacing}
\end{document}